\documentclass[aps,amssymb,amsmath,superscriptaddress,twocolumn,showpacs]{revtex4-1}
\usepackage{graphicx}
\usepackage{subfigure}
\usepackage{amsfonts}
\usepackage{dcolumn}
\usepackage{color}
\usepackage{CJK}
\usepackage{bm}
\begin{document}
\title{Expansion dynamics in a one-dimensional hard-core boson model with three-body interactions }
\author{Jie Ren}
\affiliation{Department of Physics and Jiangsu Laboratory of Advanced Functional Material, Changshu Institute of Technology, Changshu 215500, China}
\author{Yin-Zhong Wu}
\affiliation{Department of Physics, School of Mathematics and Physics, Suzhou University of Science and Technology,
Suzhou, Jiangsu 215009, People¡¯s Republic of China}

\affiliation{Department of Physics and Jiangsu Laboratory of Advanced Functional Material, Changshu Institute of Technology, Changshu 215500, China}

\author{Xue-Fen Xu}
\affiliation{Department of Fundamental Course, Wuxi Institute of Technology, Wuxi 214121, China}

\begin{abstract}
Using the adaptive time-dependent density matrix renormalization group method, we numerically investigate the expansion dynamics of bosons in a one-dimensional hard-core boson model with three-body interactions. It is found that the bosons expand ballistically with weak interaction, which are obtained by local density and the radius $R_n$. It is shown that the expansion velocity $V$, obtained from $R_n=Vt$, is dependent on the number of bosons. As a prominent result, the expansion velocity decreases with the enhancement of three-body interaction. We further study the dynamics of the system, which quenches from the ground state with two-thirds filling, the results indicate the expansion is also ballistic in the gapless phase regime. It could help us detect the phase transition in the system.
\end{abstract}
\maketitle

\section*{introduction}
\label{sec:introduction}
Recently, the understanding of nonequilibrium dynamics of strongly correlated many-body
systems poses one of the most challenging problems
for both theoretical and experimental physics\cite{Bloch}.
Researches in the nonequilibrium properties of strongly
correlated many-body systems have emerged into a dynamic
and active field, driven by the possibility to address questions
such as thermalization \cite{Polkovnikov,Rigol,Sedlmayr} and particle transportation
\cite{Ott,Fertig,Strohmaier,Schneider,Rigol2004,Rigol2005,Tyer,Minguzzi,delCampo} in clean, well-controlled, and isolated systems. In the particle transport cases, there are two prototypical transport mechanisms in classical physics: ballistic and diffusive transports. The ballistic and diffusive transports are characterized by non-decaying currents and decaying currents respectively. In the ballistic systems there is absence of friction. However, in the diffusive systems, there have frequent diffractive collisions, which drive a local thermalization. Many studies
focus on qualitative questions such as whether transport is ballistic or
rather diffusive in microscopic models of strongly interacting
systems\cite{langer2009,Langer2011,langer2012,Vidmar,Heidrich,Boschi}. In experiment, the measurements of local occupancy dynamics can be realized even the densities of initial states are than one\cite{Xia}. The dynamics exhibit clear signatures of quantum distillation and confinement of
vacancies in the doublon sea, which are significantly interesting. Other typical examples investigated numerically are the expansion of initially localized ultra-cold bosons in
homogeneous one- and two-dimensional optical lattices. It is found that both dimensionality and interaction
strength crucially influence these nonequilibrium dynamics, which have also been confirmed in experiments\cite{Ronzheimer}.

Previous results for nonequilibrium dynamics strongly correlated many-body
systems are almost with the dominant two-body interactions, because relatively small
multi-body interactions can only provide tiny corrections. Recently, the cold atom in optical lattice gives us a great
platform to realize three-body interactions \cite{Zoller,Will,Petrov}. It is shown
that the three-body interactions can be dominated, and the
two-body interactions can be independently controlled and even
switched off by driving microwave fields\cite{Zoller}. The system
with multi-body interactions can induce many exotic
phenomena. It would be interesting to investigate nonequilibrium dynamics in the system with multi-body interactions, such as
sudden expansion of Mott insulators (MI) in a one-dimensional hard-core boson model with three-body interactions.

The paper is organized in the  following ways. We define the
model Hamiltonian, and the observable is also provided. The evidence of the difference between ballistic
and diffusive expansion dynamics, which can be obtained by the behaviours of the observable, is displayed. Then we focus on the sudden expansion
in a one-dimensional hard-core boson model with the three-body interaction. The results presented in the main text are obtained
for the expansion from the box trap. The hard-core bosons that expand from the product of local MI states is probed. Furthermore, we investigate the hard-core bosons expansion from the entanglement states. A discussion follows in the last section.

As a result, when the hard-core bosons expand from the product MI states, we obtain the bosons expansion ballistically with weak interaction. The expansion velocity decreases with the increase of three-body interaction.  Moreover, we study the dynamics in the system from the ground state with two-thirds filling, and our results indicate that the expansion is also ballistic in the gapless regime.

\begin{figure*}
\centering
\includegraphics[width=0.9\textwidth]{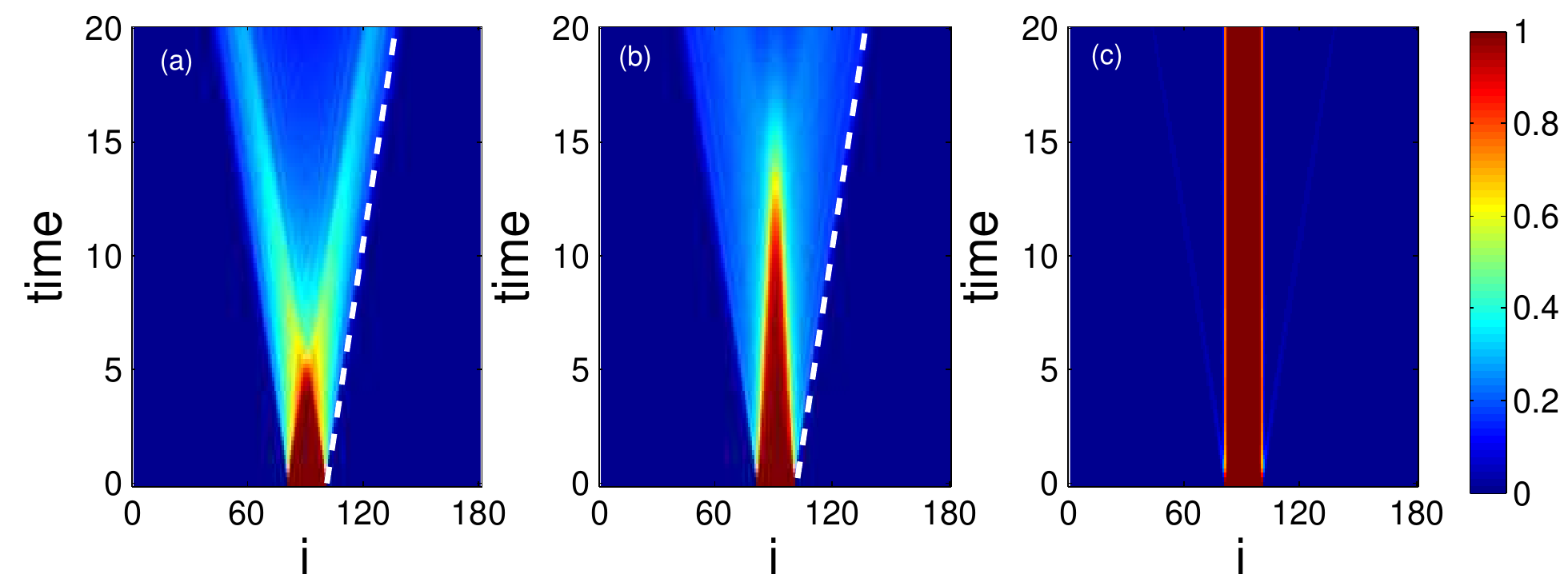}
\caption{(Color online) Typical contour plot of the local density $ \langle n_i(t)\rangle$ as function of position and time during the expansion from a MI(N=20,L=180) with different three-body interaction (a)$W=0$,(b)$W=1$ and (c)$W=3$. The slanted lines in (a) and (b) indicate the speed 2J. It is noted that the strength of (a),(b) and (c) has the same colour-bar.}
\label{fig1}
\end{figure*}
\section*{ Model and  Measurements }
\label{sec:model and  Measurements}

The Hamiltonian we consider is one-dimensional hard-core
boson model with three-body interactions, and is given by
\begin{eqnarray}
\label{eq1}
&H=-J
\displaystyle{\sum_i}(b_i^\dagger b_{i+1}+b_ib^\dagger_{i+1})+W\displaystyle{\sum_i}n_{i-1} n_i n_{i+1},
\end{eqnarray}
where $b^\dagger_i(b_i)$ is creation (annihilation) operator of
hard-core boson at site $i$: $(b^\dagger_i)^2=(b_i)^2=0$, and $n_i=b^\dagger_ib_i$ is local density operator. The parameter $J$ is the hopping interaction and
chosen as the unit of energy in the paper, and only the leading three-body
interactions with strength $W$ is considered. Open boundary conditions are imposed in the system.

As we known, there are two prototypical transport mechanisms in classical physics: ballistic and diffusive transports. The two kinds of transports can be distinguished by using the time dependent radius of the density distribution, which is defined as
\begin{eqnarray}
\label{eq2}
R_n(t)=\sqrt{\frac{1}{N}\sum_{i=1}^L \langle n_i(t) \rangle (i-i_0)^2-R^2_n(t=0)},
\end{eqnarray}
where the parameter $N$ indicates the number of the particles and $L$ indicates the length of lattice sites. The parameter $i_0$ represents the center of mass $i_0=L/2+1/2$. A ballistic expansion will lead to the well-known
$R_n(t) \propto t$ behavior, while a diffusive expansion
with a fixed diffusion constant, will lead to the well-known
$R_n(t) \propto \sqrt{t}$ behavior. The radius has the ability to detect transport whether if is ballistic or diffusive.  These have also been verified for spin and energy
dynamics in the spin-1/2 XXZ chain\cite{langer2009,Langer2011}.

Several powerful methods have been employed to
study the expansion dynamics in one dimension, including
the time-dependent density matrix renormalization group
method (t-DMRG)\cite{U2005,Vidal,white,U0211,Daley}. The method  is also applied in the paper. In our simulations, the Trotter slicing $Jt= 0.1$ is used, and the t-DMRG codes with double precision
are performed with a truncated Hilbert space of $m= 400$. The time that can be simulated in the system is determined by the entanglement entropy.

\begin{figure}[t]
\includegraphics[width=0.45\textwidth]{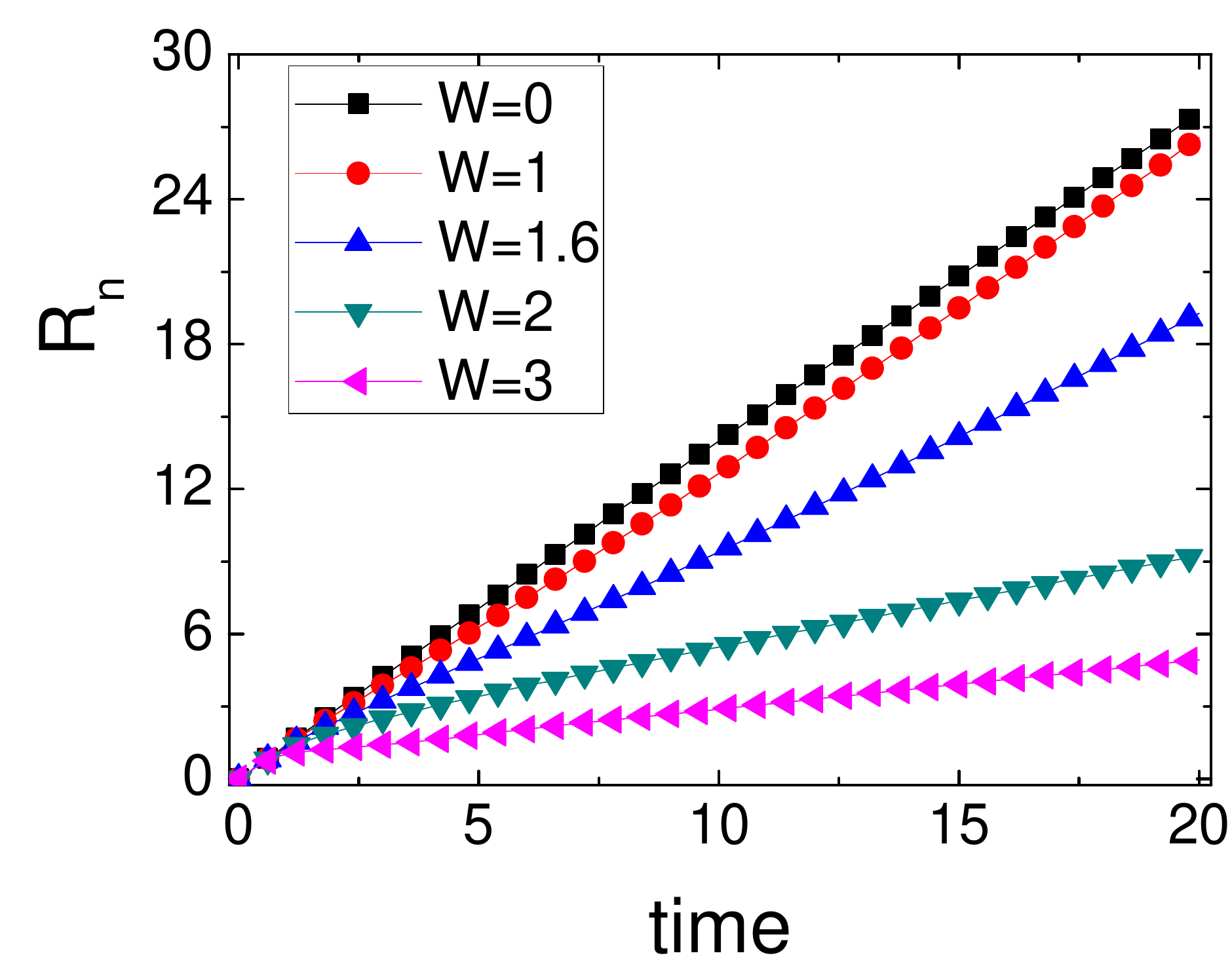}
 \caption{(Color online) The radius $R_n(t)$ is  plotted as function of time for different W  with $N=20$.}
\label{fig2}
\end{figure}

\begin{figure}[t]
\includegraphics[width=0.5\textwidth]{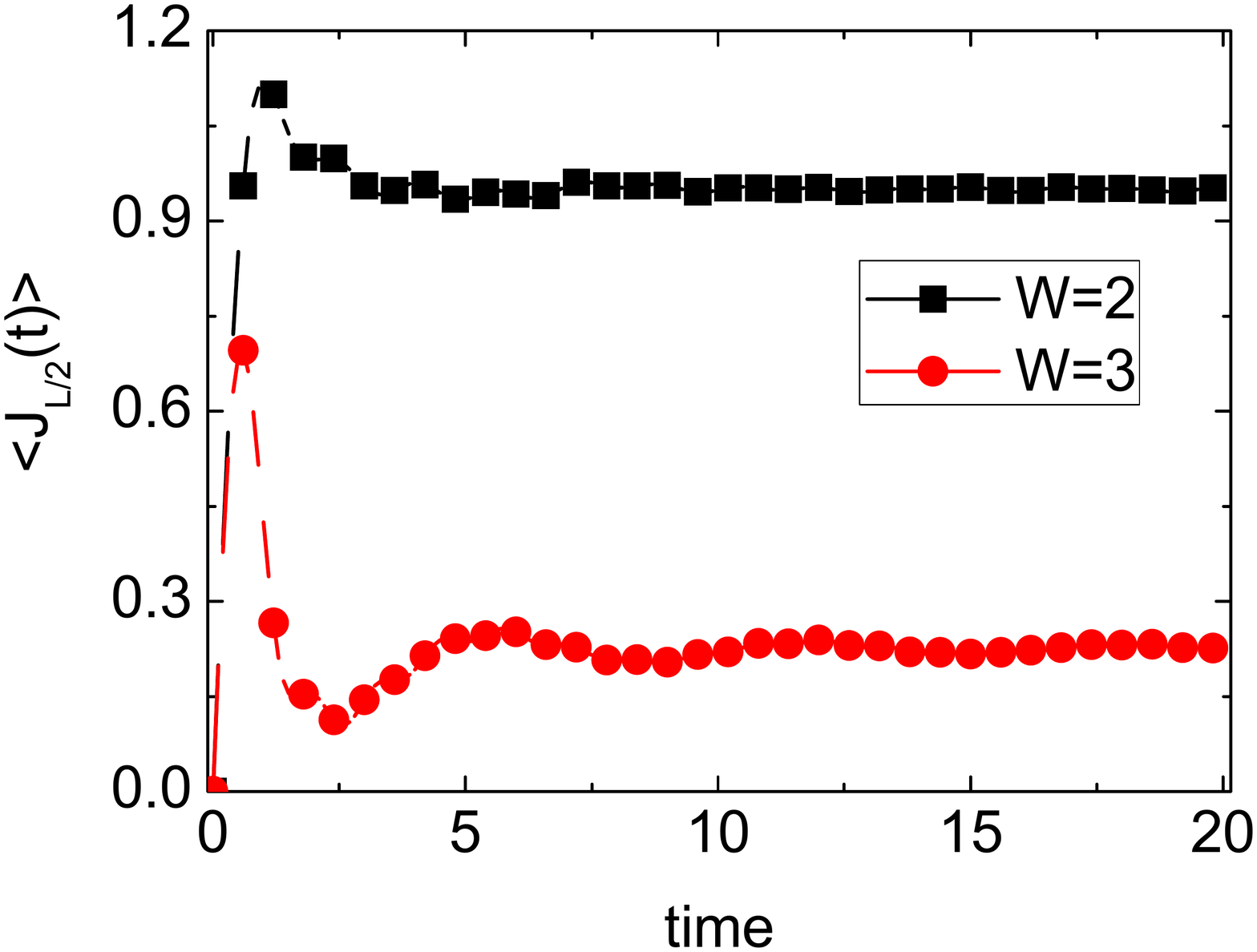}
\caption{(Color online)The total particle current $J_{L/2}$ vs $time$ for different three-body interaction $W$. }
\label{fig3}
\end{figure}

\begin{figure}[t]
\includegraphics[width=0.45\textwidth]{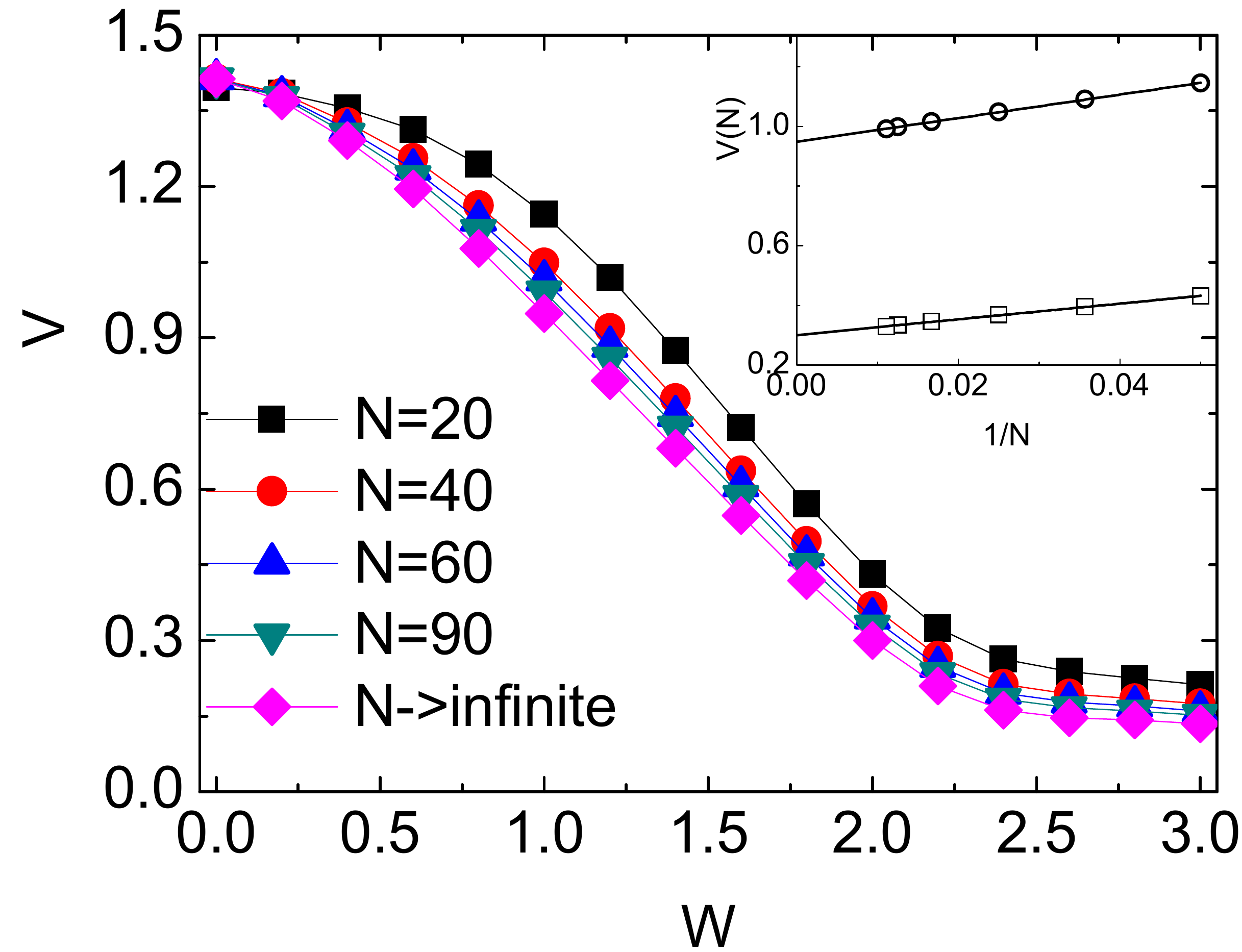}
\caption{(Color online) The expansion velocity $V$ vs $W$ with different N. The inset: Finite-size scaling of $V(N)$ with $W=1(\circ)$ and $W=2(\square)$. The lines are the fit lines.}
\label{fig4}
\end{figure}

\section*{Expansion from MI state}
\label{sec:Fock state}
For  simplicity, we first discuss the idealized case that all the particles are in a box-like trap. It is not difficult to achieve in experiments, just like by using the confining potential \cite{Vidmar}.
It is important to emphasize that the confining potential would influence the three-body interactions. After relaxation of the confining potential, its effects on the three-body interaction will disappear. As is known, the theoretical investigation for two-body interactions in such a system is very interesting, and has been investigated in Refs.\cite{Gobert,Karrasch}. Here, we will focus our attention on the effects of three-body interactions. The initial states are that all the particles are in a box trap states ($L_{box}$ is the length of the box) and the box trap is full of with particles, and the initial state is a product of local MI state in the box trap, and is given by
\begin{eqnarray}
 \label{eq3}
|\Psi_{initial}\rangle=\prod_{i_1 \leq i \leq i_2} b_i^\dag|0\rangle.
\end{eqnarray}
The parameter $N$ indicates the number of the particles $N=L_{box}=i_2-i_1+1$, and this state has been realized in a recent experiment\cite{Ronzheimer}.

A typical example for the time evolution of the density $\langle n_i(t)\rangle$ is
shown in Fig. \ref{fig1} for the expansion from a MI state with different W.
In Fig. \ref{fig1}(a), it is found that the MI melts on a time scale of $t_{melt} \simeq L_{box}/2/(2J)=5$ for $W=0$. For the case of $W=0$, the Hamiltonian (\ref{eq1}) can be diagonalized by using Jordan-Wigner transformation and Fourier transformation, which is given by
\begin{eqnarray}
\label{eq4}
H=\sum_k \epsilon_k c^\dag_k c_k,
\end{eqnarray}
where
\begin{eqnarray}
\label{eq5}
\epsilon_k=-2J cos(k),k=l\pi/(L+1),l=1,2,3\cdots L.
\end{eqnarray}
Therefore $2J$ is the largest possible velocity. In Fig. \ref{fig1} (a) and (b) two lines are parallel to the outer rays visible in the figure,
i.e., the fastest propagating particles, and it indicates an excitation spreading out from the center box with the same group velocity  $2J$. When $W=1$, $t_{melt}$ increases, as shown in Fig. \ref{fig1}(b).
For $t > t_{melt}$, two particle clouds form that propagate into opposite directions,
visible as two intense jets, as is expected for ballistic dynamics in one dimension system\cite{langer2009,Langer2011}. When $W=3$, we cannot find  $t_{melt}$ in  Fig. \ref{fig1} (c). With strong three-body interaction, the particles are almost all confined in the box trap. There is no possibility of compensating the loss in interaction energy with kinetic energy in the system due to energy conservation.

In order to clearly distinguish whether the expansion is ballistic or not, we also investigate the radius behaviors. In Fig. \ref{fig2}, we show the radius $R_n(t)$ for different magnitude of three-body interactions. Clearly when $W=0,1.0,1.6$, the radius $R_n(t)\sim t$ within short and intermediate time. In the time region, the local densities are large, and the expansion could be considered ballistic.
For $W=2.0, 3.0 $, in the initial time it does not follow  $R_n(t)\sim t$, and after that the radius follows $R_n(t)\sim t$. It seems that the expansion is ballistic obtained by the radius. But in general, ballistic dynamics in strongly interacting dynamics is connected to integrability. The hard-core bosons without any additional three-body interactions expand ballistically due to their integrability
and the mapping to non-interacting fermions that preserves density. The three-body interactions break integrability
and the dynamics would be expected as diffusion. With strong three-body interactions, that the particles are almost all confined in the box trap also suggests that the transport would be diffusive\cite{Polini}. However, the sudden expansion of initially trapped particles into an empty lattice is more complicated, this can not provide sufficient evidence for diffusion. To be sure, the intermediate time regime in which $R_n$ is not proportional with time could be an indicator of non-ballistic dynamics.

In order to check the results for $W=2.0,3.0$, we investigate the time dependence of the total
particle current in each half of the system, which is defined as
\begin{eqnarray}
\label{eq6}
J_{L/2}=-i\sum_{i>L/2}J(b_i^\dagger b_{i+1}-h.c.).
\end{eqnarray}
It is noted that ballistic dynamics are probed into the systems
with open boundary conditions  due to the existence
of globally conserved currents\cite{Rigol2008}.
The results of $J_{L/2}$ for $W=2,3$ are shown in Fig. \ref{fig3}. The parameter $J_{L/2}$ takes a constant value for $W=2,3$, which is the
benchmark feature of  ballistic dynamics\cite{Langer2011}.  However, the behaviour of $J_{L/2}$ does not imply that the dynamics will be ballistic. There is a subtle solution of the equation that the radius increases linearly with time (if properly defined in Eq. (\ref{eq2})), and the total particle current in each half of the system takes a constant. The phenomena generally occur in the dilute limit. With strong interaction $W=2,3$, the particles are almost all confined in the box trap, and only a few particles transfer. This case could be considered the dilute case, and the dynamics will expand ballistically at a very long time, irrespective of whether it is ballistic or diffusive mass transport.

We extract the expansion velocity
$V$  by fitting the tDMRG data , i.e., the slope of curves such as the
ones shown in Fig. \ref{fig2}, with Curve Fitting toolbox in Matlab. The corresponding radial
velocity $V$ is defined through the reduced radius $V={\partial R_n(t)}/{\partial t}$. As is known,
there are two main sources of errors in the adaptive t-DMRG: the Trotter error due
to the Trotter decomposition and the DMRG truncation error due to the representation of
the time-evolving quantum state in reduced Hilbert spaces.
For small times, the Trotter error plays a major role. For long times, the DMRG truncation error will be dominated\cite{Daley,Ren}. We obtain $V$ with the date $R_n(t)$ in time range $2 \leq t\leq10$, where both Trotter error and truncation error are small. Results for selected values of $W$ are
collected in the main panel  of Fig. \ref{fig4}. For the expansion from the MI
we obtain $V = \sqrt{2}J$  for any $N >0$ and $W=0$ \cite{Boschi}. The expansion velocity also decreases with  the increase of $W$  for the same $N$. This results in an increase of interaction energy and therefore a decrease in kinetic energy. The velocity $V$, which embodies the behaviours of kinetic energy, decreases when the interaction increases. The expansion velocity is  much smaller than the largest possible velocity $2J$ and is always very different from characteristic velocities of the initial state.
Fig. \ref{fig4} also indicates that the expansion velocity decreases while $N$ increases, and the finite-size effect is shown in the inset. For above case, the length of system does not affect the propagation noticeably until the particles wave front reaches the boundary. We can fit the locations of maximums by the formula
\begin{eqnarray}
\label{eq7}
V(N)\sim V_c+aN^{-1},
\end{eqnarray}
where $a$  is a size-independent constant and $N$ the number of bosons\cite{Vidmar}. We obtain that $V_{c1}=0.95$ for $W=1$ and $V_{c2}=0.30$ for $W=2$,  see Fig. \ref{fig4} inset. The results $N\rightarrow\infty$ for different $W$ is also shown in Fig \ref{fig4}.  The above results about the expansion velocity can provide some information about the initial state.

\begin{figure}[t]
\includegraphics[width=0.45\textwidth]{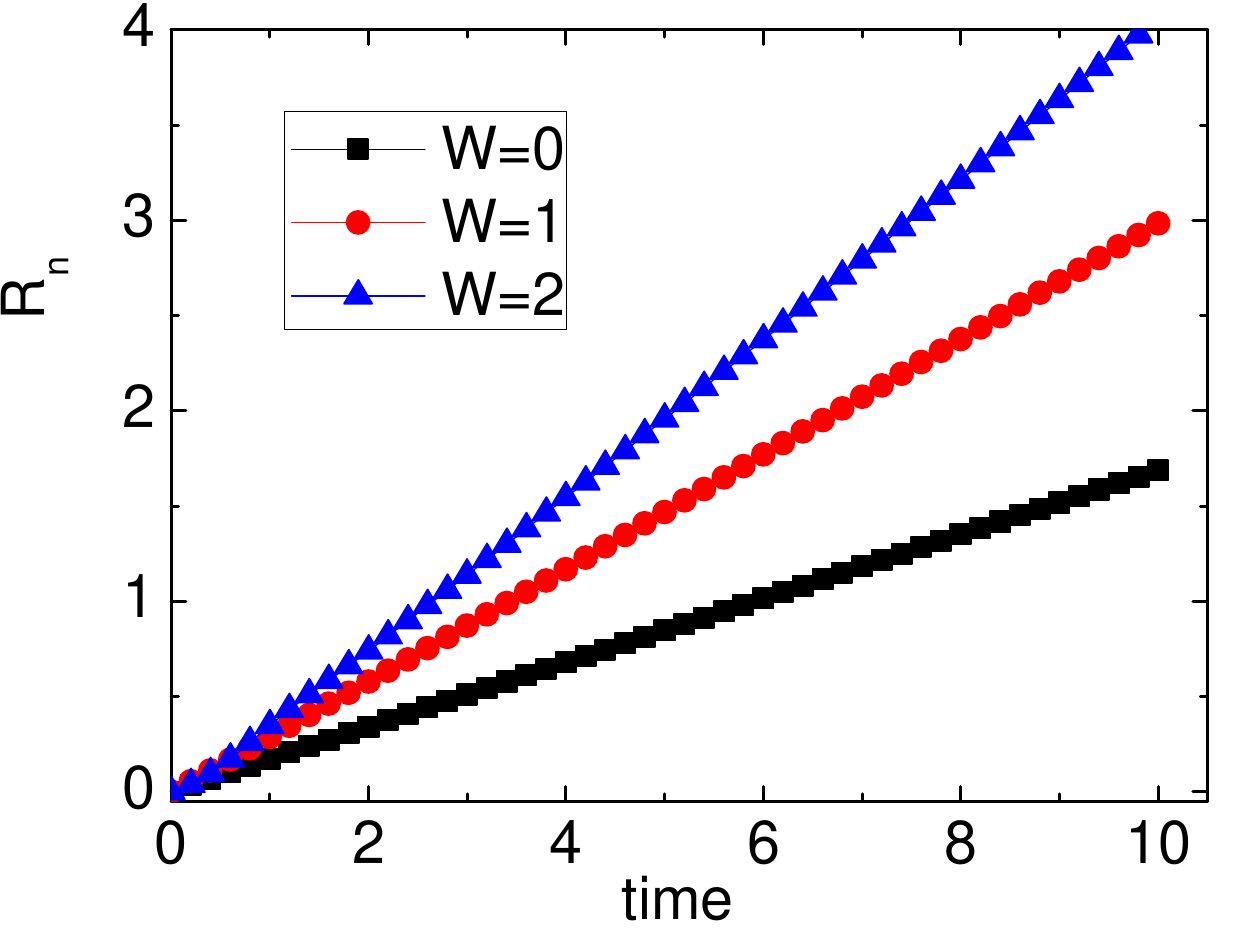}
\caption{(Color online) The radius $R_n(t)$ is plotted as function of time for different W with  $\langle n\rangle=N/L=2/3$.}
\label{fig6}
\end{figure}

\begin{figure}[t]
\includegraphics[width=0.45\textwidth]{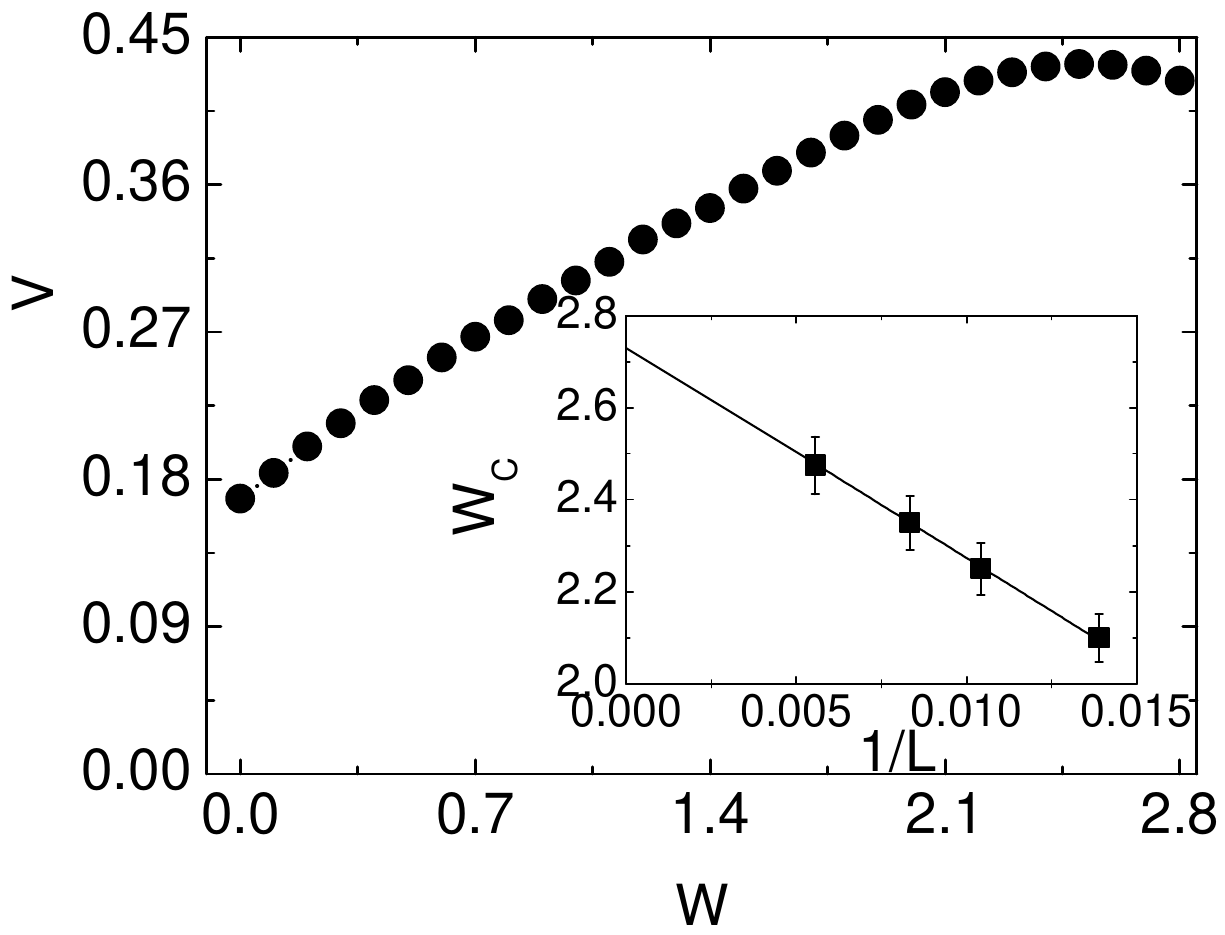}
 \caption{(Color online) The expansion velocity $V$ is plotted as function of the three-body interaction $W$ with $L=180$. Inset: Finite-size scaling of critical $V_c(N)$. The lines are the fit lines.}
\label{fig7}
\end{figure}

\section*{Expansion from the grand-state}
\label{sec:ground state}

In this section, the initial state can be realized by finding the ground
state of a system with $(N-2)$ and then applying
two operator flips $b^\dag_{L/2}b^\dag_{L/2+1}$. The time evolution is then performed
at the two-thirds filling $\langle n\rangle=N/L=2/3$. It is noted that with this filling, the quantum phase transition from the superfluid to solid phases in the system is located at $W_c \simeq 2.80\pm0.15$, which are obtained by the structure factors and bond order parameter\cite{Sansone}. In order to find whether the expansion is ballistic or diffusive, we also investigate the radius behavior. In Fig. \ref{fig6}, we display the radius $R_n(t)$ for various three-body interactions for $\langle n\rangle =2/3$. Clearly when $W=0,1.0,2.0$, the radius is proportional with time$R_n(t)\sim t$. It seems that the expansion is ballistic.
It is noted that the radius obtained by Eq. (\ref{eq2}) does not follow $\propto t$ or $ \propto\sqrt{t}$ when $W\geq2.8$ due to the strong oscillations of $\langle n_i(t) \rangle$ \cite{langer2009}.  We also adopt the method of Eq. (13) in Ref. \cite{langer2009}, the oscillations also exist. We cannot conclude the transport is ballistic or diffusive by the radius. We extract the expansion velocity $V$  by fitting the tDMRG data. We obtain $V$ with the data $R_n(t)$ in time scale $2 \leq t\leq10$, which the Trotter error and truncation error are both small. The velocity for different values of $W$ are collected in the main panel of Fig. \ref{fig7}. It is seen that the velocity has a peak which is located at $W\simeq 2.5$. The initial states would be influenced by the finite size effect. The location of peak may change for different system size. So we investigate the location of peak with different system sizes. It is found that the locations of expansion velocity maximums follow the formula
\begin{eqnarray}
\label{eq9}
W(L)\sim W_c+cL^{-1},
\end{eqnarray}
where $c$  is a size-independent constant and $L$ the system size, seen in Fig. \ref{fig7} inset. It is found that the critical point $W_c \simeq 2.74\pm0.1$, which is close to the the superfluid to solid phases quantum phase transition point. The dynamics would be diffusive when $W>W_c$\cite{M,Prosen,Benenti}, where the system is in the gapped phase\cite{Luo}. It means that the peak of velocity will move to the boundary between two kinds of transport behaviors, as the system size approaches to infinite.

\section*{Discussion}
\label{sec:Discussion}

In the paper, we study the expansion Mott insulator in a one-dimensional hard-core boson model with three-body interactions by using the adaptive time-dependent density matrix renormalization group method. We obtain that the bosons expand ballistically with weak interaction by studying the dynamics of local density and the radius $R_n$. It is also found that the expansion velocity $V$, is dependent on the number of bosons in the system, and its measurement can provide information about the initial state. As a result, the expansion velocity decreases when the strength of three-body interaction increases. Moreover, we study the dynamics of the system from the state with the two-thirds filling. Our results indicate that the expansion also is ballistic in the gapless phase. In the case of the two-thirds filling, the superfluid to solid phases quantum phase transition point can be identified by the peak of the expansion velocity. It would be interesting to test our predictions in experiments. With strong interaction, the dynamics could not be ballistic. It would be significance work to get the expansion dynamics by other parameters.

\section*{acknowledgments}
We thank Shi-qun Zhu and Xiao-qun Wang for valuable discussions.  This work is supported by the National Natural Science Foundation of China (NSFC) (Grant Nos 11174114, 11274054).

\section*{Author Contribution}

J.R. conceived the idea. J.R. and Y.Z.W. performed the simulations. J.R. and X.F.X. analyzed the results of simulations. All authors co-wrote the manuscript, and contributed to discussion and reviewed the manuscript.

\section*{Additional Information}

\textbf{Competing financial interests:} The authors declare no competing financial interests.\\

\end{document}